\documentclass[prd,preprint,eqsecnum,amsfonts,amssymb,amsmath,floatfix,nofootinbib]{revtex4}

\usepackage{bm}
\usepackage{slashed}

\begin{document}

\begin{flushright}
BRX-TH-6289             \\
CALT-TH-2015-003
\end{flushright}

\title{The legacy of ADM}

\author{S. Deser} 
\affiliation{Walter Burke Institute for Theoretical Physics, Caltech, Pasadena, California 91125, USA \\ 
Physics Department, Brandeis University, Waltham, Massachusetts 02454, USA}

\begin{abstract}
This tribute to the memory of my old
friend and collaborator, Richard Arnowitt,
focuses on the history, results and
physical significance of the Arnowitt-Deser-Misner (ADM) formulation of General
Relativity, starting from its birth in 1958-9
through its completion, in a series of over
a dozen papers, in 1962-3. A few of its
later applications are also mentioned.
\end{abstract}

\maketitle

\section{\label{sec:sec1}Prehistory}

Dick and I came to Harvard's Physics
graduate school essentially
simultaneously, although he had already
earned an MA from RPI.
Our paths were parallel 
throughout Harvard's course system for the first two
years and when we both began
theoretical research under our mentor,
Julian Schwinger, then an almost
unapproachable idol. Those were the years right after the wave
of wartime students' influx
(Julian alone producing many Nobel
laureates, besides himself), and marked
the end of the great Quantum Electrodynamics (QED) postwar era.

We each had different postdoc 
trajectories, but in Fall '58 I found myself
back at Harvard as Schwinger's assistant,
at a time when
everyone was trying to find the next wave.
Although I worked on the then fashionable
dispersion theory, both Dick and I
maintained our love for Quantum Field Theory. Indeed, so
did Julian, who began to (re-)think about
higher spins, including spin 2 -- the basis
of General Relativity (GR), which he of course
had learned as a child. At that time, GR
was an almost forgotten domain, mostly
devoted to the botany of exact solutions
and other geometrical practices, and
frowned on as an occupation for young
theoreticians by that era's authorities.
[What a difference a few decades makes!]
Independently of Schwinger, Dick and I
began to apply our knowledge of QED to
spin 2 as well, and in that year ’58-’59,
had completed the canonical analysis of
the easy part -- linearized GR. It goes
without saying that, for us quantum
dwellers, the aim was to quantize the
theory. [Of course we knew that
quantizing the linear limit would be trivial -- as indeed it was, having been done in
the early thirties by Bronstein and Rosenfeld-- and that perturbative
GR was non-renormalizable, by the
already standard old Fermi theory
argument using the dimensionality of
Newton's constant.]

During that first period, we were
summoned by Wheeler, who had heard
of, and wanted to be briefed on, our
ideas. As we entered his office -- which
was equipped with a huge tape recorder
as well as his notorious bound notebooks -- he asked whether he could bring in his
student who might have some relevant
results for our quest. Being seasoned
postdocs, we did not expect much from a
mere student, but agreed. After our
exposition, Charlie Misner told us of his
new formulation of the GR action, and it
was suggested we join forces. We were
smart enough to realize the benefits and
rapidly agreed to do so. Thus was ADM
born -- and none of us ever looked back.

\section{\label{sec:sec2}ADM: the history}

Between 1958-9 and 1963, the three of us
collaborated, sometimes in the same
spot(s) -- they were many -- by mail, and
(rarely) by phone in those prehistoric
days. We attended the historic
Royaumont Conference in France,the
third in the famous series (the traditional
numbering is \#0 in Bern in 1955, which
only I attended, as an ignorant tourist, \#1
was in Chapel Hill in 1957 where Charlie
\& I were present). In Royaumont, we
discovered that Dirac was on the same
track -- but also that our approaches were
almost orthogonal, and that we were
ahead in results: This was fortunate,
because to compete with Dirac would
otherwise be enough to discourage
anyone! The whole subject was indeed
slowly waking from its hibernation from
the early twenties: for example, even that
first dedicated, 1955 Bern, conference
had but 80 participants, of whom far fewer
were pros. Still, by the late ‘50s there
were a number of active, if small, groups
of younger people working in London,
Syracuse, Texas, UNC, Hamburg,
Warsaw, Princeton and the USSR. The
Warsaw conference of 1962 also included
Dirac, as well as Feynman -- who mainly
(co-) contributed the formal need for
ghosts in covariant quantization of nonabelian
gauge fields.

The most intense, initial -- and productive -- 3-way stretch came in the Summer of '59,
when we all met on a Danish island and
worked on the floor in a Kindergarten,
now alas gone, on knee-high
blackboards. It was there that we found
the basic translation of the geometrical
meaning of GR into modern Field
Theoretical language. From that
realization, all the pieces fell into place -- the many results we obtained in our
many subsequent papers just kept
coming! By 1963,we had more or less
found all that was worth doing to set up
the physics, and our collaboration -- but
not our friendship -- naturally dissolved on
geographic grounds.

\section{\label{sec:sec3}ADM -- the physics}

Since this is intended for a nonspecialized
readership, I will only outline
the basics of the ADM version of GR;
details and extensions may be found in
our various papers; the entire story,
including individual references, is
contained in our major 1962 review paper,
now available online~\cite{deser1}.
I emphasize that our formulation is
entirely equivalent to geometrical GR, but
couched in the then newly developed language of gauge theories. The key equation from which
many results flow is the Einstein-Hilbert
action (the densities also agree up to a total divergence) in $3+1$ [or in any $(D-1)+1$], in canonical,
Hamiltonian -- $L=p \dot{q} - H(p,q)$ -- first
order form and in Planck units,
\begin{eqnarray}
I &=& \int d^4 x \left [\pi^{ij} \dot{g}_{ij} - N_\mu R^{\mu} \left( \pi,g \right) \right]
\label{eq1a} \\ 
R^0 &=& \sqrt{ ^{3} g } \; ^{3} R +\left( \pi^2 / 2- \pi^{ij} \pi_{ij} \right) / \sqrt{ ^{3} g } 
\label{eq1b}  \\
R^i &=& -2 D_j  \pi^{ij} \; .
\label{eq1c}
\end{eqnarray}

Here the six conjugate pairs ($\pi^{ij}, g_{ij}$) are
of course to be varied independently and
all operations such as index-shifting and covariant differentiation $D_i$ use
the 3-metric $g_{ij}$ and its inverse $g^{ij}$.
All variables are combinations of the
metric and second fundamental form:
\begin{eqnarray}
\pi^{ij} &\equiv & \sqrt{-g } \, g^{ip} g^{jq} \left (\Gamma^0 _{pq} - g_{pq}
\Gamma^0 _{rs} g^{rs} \right) 
\label{eq2a} \\
N_0 &\equiv & N \equiv \left ( -g_{00} \right) ^{-1/2} , N_i \equiv g_{0i} \; .
\label{eq2b}
\end{eqnarray}

So $\pi^{ij}$ is essentially the time derivative of
the spatial metric $g_{ij}$, as befits a
canonical momentum variable, while the
Lagrange multipliers $N_{\mu}$ in (\ref{eq1a}) are
collectively the set \{$g_{0 \mu }$\}, nowadays also
called the shift ($N_i $) and lapse ($N$).
The Hamiltonian action (\ref{eq1a}) is both very
familiar (as $p \dot{q} - H$) and very
alien: varying the $N_{\mu}$ yields the four
constraints $R^{\mu}=0$ -- constraints because
they do not involve time derivatives -- leaving a vanishing Hamiltonian: now $L=
p \dot{q} - 0$ !

The resolution of this paradox is key to
understanding how geometry,
including coordinate invariance,
transmutes into ``mechanics''.
What we realized in that Kindergarden
was that this phenomenon was already -- if artificially -- constructed long ago by
Jacobi -- who showed that the standard
action principle for a dynamical system
could be parametrized to elevate the
number of degrees of freedom by one,
and remove the Hamiltonian altogether,
replacing it by a constraint on the that
new -- fake -- excitation:
\begin{eqnarray}
I &=& \int L d\tau \\
&=& \int \left( \left[ \sum\limits_{a=1}^N p^a dq_a + P dQ \right] -N \left[ P + H(p,q) \right] d\tau \right) \\
&=& \int \left( \sum\limits_{a=1}^{N+1} p^a dq_a - N R \left( P; p,q \right) d\tau \right) \; .
\label{eq3}
\end{eqnarray}

Solving the $R$-constraint for $P$ and choosing
$Q$ to be the time recovers the original
action. Note the ``general covariance'' of
the last expression under choice of time $\tau $, and
its exact ADM form (\ref{eq1a}). So GR is what we
called an ``already parametrized'' theory -- with no underlying ``normal'' action $I= \int dt
\left[ p \dot{q} - H(p,q) \right] $ and time $t$ to take refuge
in! There are four constraints, simply
because we have a field theory with four
coordinates instead of just time. 
Each constraint removes one degree of freedom (in this first order form), leaving only two of the original 6 
$\left( \pi^{ij}, g_{ij} \right) $ pairs. Concretely, we may take these to be the two transverse-traceless (TT) modes (just two because of the 4 conditions $\partial_j \pi^{ij}= 0= \pi^{i}_{i} $, etc.). [More generally, they are the equivalents of the two transverse photon excitations $\left( E^{i \,T}, A_{i}^{T} \right) $ -- and indeed of the two helicity $\pm s$ modes of any massless, spin $s >0$ excitations, in $D=4$.] In the weak field limit, they describe the -- abelian gauge-invariant -- helicity $\pm 2$ gravitons. We are separately 
free to decide on what to use as the
coordinates $x^{\mu}$, including time -- this is
coordinate invariance of course -- with the
corresponding, conjugate, choice of
Hamiltonian or rather 4-momentum $P_\mu$.
[This gauge freedom comes with all sorts
of problems of principle regarding gauges
that differ by functions of the two ``true''
excitations, especially in any formal
quantization attempt. We discussed
these troubling issues in some detail,
but they keep being rediscovered.]
The physical guide to proper coordinate
choices, apart from more exotic
questions for spaces with weird topology, is of course the asymptotic -- at spatial
infinity (rather than null infinity, more
suited to non-``3+1'' approaches) -- weak
field regime, that is, the usual choice of
boundary conditions in field theories,
though here there can also exist closed
spaces with no infinity -- and hence no
notion of energy either, as we will see.
[Parenthetically, a cosmological constant
is easily incorporated, since $\sqrt{-g} =N \sqrt{ ^{3} g } $, by just 
adding $\sqrt{ ^{3} g }$ to $R^0$ in (\ref{eq1a})
and altering asymptotic states from flat to
(A)dS spaces. 

Gauge invariance means
local quantities such as the gravitational
field's stress-tensor are meaningless -- but
global ones such as $P_{\mu}$ are physical -- in
particular, energy is well-defined, and is in
fact even simply expressed as a surface
integral over energy flux at spatial infinity,
just like flux of longitudinal electric field
counts the total charge $Q$ in Maxwell. This
so-called ``ADM'' energy or mass plays an
essential role in every aspect of GR, and
indeed in Supergravity (SUGRA), GR's modern
successor~\cite{deser2}.
In its simplest terms, $E$ is defined from
the $R^0$ constraint of (\ref{eq1a}), as the conjugate
to the simplest asymptotic time choice,
best seen by expanding $R^0$ into its
unique linear term -- the linear part of the
3-curvature $^{3}R \sim \nabla ^{2} g^T$, where $g^T$ is
a particular component of $g_{ij}$ in its
orthogonal expansion that extends a
vector's decomposition into transverse
(divergence-less) and longitudinal (curlfree)
parts -- plus the nonlinear remainder.
Then the spatial integral of the Poisson
equation $\nabla ^2 g^T = - R^0 $ (nonlinear) is
the total energy. [Solutions that do not
decay sufficiently rapidly at spatial infinity,
or, at the other extreme, closed spaces,
do not -- and should not -- have well defined
$E$.] These simple arguments can
be formalized in terms of asymptotic
Killing vectors, obeying $D_{ ( i } K_{j ) } =0$;
indeed this must -- and can -- be done
explicitly to define energy also in the 
presence of the cosmological term~\cite{deser3} and
indeed for arbitrary covariant extensions
of GR~\cite{deser4}, such as $R+R^2$ models. It also
has a counterpart in the need for
asymptotic Killing ``color'' vectors to define
the total, non-abelian, charge in Yang-
Mills models~\cite{deser5} that -- unlike Maxwell -- are ``charged'', just as the gravitational
field's ``charge'' is energy.

Establishing $E$-positivity was one of the
longest-standing challenges in the field, to
which ADM contributed only some special
cases; indeed it was not proved until
much later. On the other hand, positivity
of (necessarily quantum) SUGRA energy, which includes that of
GR in the classical limit, was easily
established~\cite{deser6} soon after SUGRA itself,
basically because that theory is the ``Dirac
square root'' of GR -- the very words
``square root'' almost embody it.

Many other uses of the ADM formulation
were explored -- the wave zone and
gravitational waves could be defined, in
analogy with the near and far zones of
electrodynamics -- this being one major
example; to this very day sophisticated
numerical studies of radiation use the
ADM methods that are especially suited
to time evolution of the gravitons'
excitation modes as well as to their
creation and absorption.

Another deep problem is that of self-energy
of massive neutral or charged
particles, and more generally the effective
particle-particle
post-Newtonian interactions. The latter
subject has also been pushed to high
analytic and numerical order, using the
ADM effective matter coupling results; this
industry is now at high-$n$ post$^n $-Newtonian level, although of course,
beyond a certain order, inclusion of
radiative effects is unavoidable. Indeed it
is a beautiful (and \textit{a priori} amazing) result
that all matter systems couple to gravity in
exactly such a way as to keep the sacred form
\begin{eqnarray}
L \left( \mathrm{matt}; g_{\mu\nu} \right) 
= \Sigma_{\mathrm{matt}} p \dot{q} - 
N_{\mu} R^{\mu} \left( p,q; g_{ij} , \pi^{ij} \right) \; . 
\label{eq4}
\end{eqnarray}
That is, the coupled theory maintains the
hallmark of ``already parametrization'' -- or
general covariance -- in particular the $NR$
form, with $R$ independent of $N_{\mu}$ -- when
the conjugate matter pairs $(p,q)$ are
properly chosen. This general property
was later established in a series of long
technical papers~\cite{deser7}.

Coming back to the self-energy problem,
it had long~\cite{deser8} been speculated that GR
might dampen self-energy divergences of
matter. 
One amusing indication is the following 
argument showing that a distribution
of bare mass $m_0$ had vanishing ADM
mass $m$ in the point ($R \rightarrow 0$) limit, and
further that a charged one's limiting mass
was proportional to its charge, $m \rightarrow |e|$. It
is in fact quite intuitive: for the neutral
$m_0$, of size $R$, the ``Newtonian'' energy is
of course
\begin{eqnarray}
E(N)= m_0 - G m_0^2 / 2R \rightarrow - \infty
\label{eq5a}
\end{eqnarray}
since the self-interaction only involves the
bare, mechanical mass, whereas in GR
ALL mass *self*-gravitates, so
\begin{eqnarray}
E(GR) = m_0 - G m^2 /R \rightarrow  0 \; .
\label{eq5b}
\end{eqnarray}
The limit is obtained by solving the
quadratic equation (\ref{eq5b}) for
$m(m_0 , R)$. Physically, this means that the
total mass $m$ diminishes from its dilute
value $m_0$ at $R = \infty$. Once it reaches $0$,
however, the process stops -- $m$ cannot
go negative, there being no mass left to
gravitate!

The above amazing -- perhaps
counterintuitive -- result even extends to a
charged massive distribution: the effect of
the Coulomb interaction, namely adding
the extra term $+e^2 / 2R$ in (\ref{eq5a}) or (\ref{eq5b}), is
as follows: in the Newtonian limit, one
gets $E(N) \rightarrow \pm \infty$ depending on the ratio
of $e/ m_0 $. Instead, the GR limit is finite:
\begin{eqnarray}
E(GR) \rightarrow \sqrt{  e^2 / G } \; . 
\label{eq5c}
\end{eqnarray}

These elementary arguments are borne
out by solving the corresponding constraint equations in detail. To be sure,
classical finiteness does not imply quantum 
finiteness, although there has been some
quantum corroboration~\cite{deser9} of our above entirely nonperturbative
effect. Indeed it is known that in perturbative (our only current general 
approach) quantum gravity, even low loop
orders $L$ are divergent for GR alone ($L=2$)
or coupled to matter of any spin, massive
or massless ($L=1$, except $L=3$ for $N=1$
SUGRA)~\cite{deser10}. The only remaining hope is
the very special system of maximal,
unbroken -- $N=8$ -- SUGRA, which is still
standing -- finite -- to high, 7, loop order~\cite{deser11}.

I will not enter into the immense literature
on ADM applications to many concrete
problems, ranging from cosmology -- there, amongst others, to the dark energy, dark matter and cosmic acceleration problems -- to astrophysics, in manifold ways, especially those involving gravitational radiation and matter interactions -- to gravitational model building, including extensions of GR and SUGRA~\cite{deser12} -- to numerical integration methods, to quantization. 
I must content myself with this, telegraphic and incomplete, list of topics touched by ADM ideas -- essentially all domains of GR and its generalizations. Interested readers will easily encounter specific examples, some even speaking ADM without knowing it.

\section{\label{sec:sec4}Summary}

This tribute to my deceased long-term
collaborator and friend, Dick Arnowitt,
gives an extremely compressed, and
incomplete, view of the role played by
ADM in all areas of gravitational research,
from the conceptual to the numerical. I
refer the reader to the original works,
catalogued in~\cite{deser1}, and to the vast, and
ongoing, related literature.
ADM was meant to bring GR from 1915
into the late twentieth Century -- expressing it in the language of modern
field and gauge theories -- hoping to
clarify its relation to the other three
fundamental forces on the one hand, and
to provide, in its own domain, a bridge to
numerical and analytical applications on
the other. In these tasks it has reasonably
succeeded. Our only regret is that Dick
will not share the forthcoming Einstein
Medal for ADM in this, GR's Centennial
year.

\begin{acknowledgements}
S.D. was supported
in part by NSF and DOE grants
PHY-1266107 and \#DE-SC0011632.
\end{acknowledgements}

\end{document}